\documentclass[aps,pre,showkays,twocolumn,superscriptaddress,floatfix]{revtex4}

\usepackage{graphicx}
\usepackage{color}

\usepackage{epsfig}
\usepackage{epstopdf}
\usepackage{amsfonts}
\usepackage{amssymb,amsmath}

\begin{document}

\title[Generalized Cornu spirals] {Generalized Cornu-type Spirals and their Darboux parametric deformations}

\author{Haret C. Rosu}
\email{hcr@ipicyt.edu.mx}
\affiliation{IPICYT, Instituto Potosino de Investigacion Cientifica y Tecnologica,\\
Camino a la presa San Jos\'e 2055, Col. Lomas 4a Secci\'on, 78216 San Luis Potos\'{\i}, S.L.P., Mexico}

\author{Stefan C. Mancas}
\email{mancass@erau.edu}
\affiliation{Department of Mathematics, Embry-Riddle Aeronautical University, Daytona Beach, FL 32114-3900, USA}

\author{Chun-Chung Hsieh}
\email{cchsieh@gate.sinica.edu.tw}
\affiliation{Institute of Mathematics Academia Sinica,
Nankang, Taipei 115, Taiwan, ROC}


\begin{abstract}
We generalize the Fresnel integrals and introduce a class of planar spirals ${\cal F}_n$, which contains the Cornu spiral as the case ${\cal F}_2$. Their Darboux parametric deformations are also investigated. The ${\cal F}_3$ spiral and some of its Darboux deformed counterparts are graphically illustrated.\\


\end{abstract}

{\widetext \begin{center} $\quad$ Phys. Lett. A 383, issue 23, 2692-2697 (2019)\\
$\quad$ Received: 1 December 2018, Available online: 28 May 2019\\
doi:10.1016/j.physleta.2019.05.040  \end{center}}

\vspace{2pc}

\maketitle

\section{Introduction} 

The purpose of this Letter is to introduce a class of spirals which generalizes the Cornu spiral \cite{Fowles} and to study their properties and supersymmetric deformations using an approach \cite{RMFG} based on parametric Darboux transformations. Like in supersymmetric quantum mechanics, the parameter is the constant of integration of the general Riccati solution \cite{s1,s2,s3,s4,s5}.
For this generalization, we consider the following trigonometric functions
\begin{eqnarray}\label{Fes}
& \cos \left( \frac{p}{n} z^{n} \right)~, \\
& \sin \left( \frac{p}{n} z^{n} \right)~,
\end{eqnarray}
where $p>1$ is an arbitrary positive constant, which can be naturally chosen as $p=\pi$.
The spirals, ${\cal F}_n$, are defined through the Argand plane representation, ${\cal F}_n=X_n+iY_n$, with $X_n$ and $Y_n$, as the following integrals 
\begin{eqnarray}
	{\cal C}_n(z)\equiv X_n(z) &=& \int_{0}^{z} \cos \left( \frac{p}{n} s^{n} \right) ds~, \label{Fes1}\\
	{\cal S}_n(z) \equiv Y_n(z) &=& \int_{0}^{z} \sin \left( \frac{p}{n} s^{n} \right) ds~, \label{Fes2}
\end{eqnarray}
which we call the generalized Fresnel integrals and are parametrized by the arclength of the spiral, $s$. In principle, $n$ can be any positive integer,
but here we will discard the cases $n=0$, which is the logarithmic spiral, and
$n=1$ since the spiral ${\cal F}_1$ is the circle.
For the case $n=2$ and $p=\pi$, ${\cal C}_2(z)$ and ${\cal S}_2(z)$ are the Fresnel integrals and the spiral ${\cal F}_2$ is the Cornu spiral
that were discussed in \cite{RMFG}.

We recall here that the planar curves are characterized by the relationship between the curvature and the arc length known as the Ces\`aro formula, and by the same token the Cornu spiral is defined as the curve whose curvature increases linearly with arc length. This means that the radius of curvature $\rho$ times the arc length $s$ is constant at each point of the Cornu spiral. In the case of the generalized Cornu spirals, the Ces\`aro equation has the form $\rho (s) = s^{1-n}/p$, which for $n=2$ and $p=\pi$ reduces
to the standard Ces\`aro equation for the common Cornu spiral. 
For $n=1$ one obtains the constant radius of curvature of the circle, and for $n=0$, the radius of curvature is proportional
to the arc length which yields to the logarithmic spiral.

In this Letter, we consider the  set of generalized spirals ${\cal F}_n$
by means of the complex parameter which appears in the general solution of the Riccati equation that corresponds to the associated generalized Fresnel integrals. In Section II, we show the reduction of the third order ordinary differential equation (ODE) satisfied
by these integrals as particular solutions to the corresponding Riccati equation, whose general solution is obtained explicitly. We then write the solution of the third order ODE based on the general Riccati solution and present Argand plots of this solution. In Section III, the similarity with supersymmetric quantum mechanics is emphasized by means of the factorization approach
\cite{s1,s2,s3,s4,s5} which is applied to the second order linear ODE that comes into play in the reduction process of Section II.

\section{The ${\cal F}_n$ spirals: The third and second order ODEs and the Riccati equation}
The following linear third order ODE
\begin{equation}\label{eq1}
z{\rm w}'''-(n-1){\rm w}''+p^2 z^{2n-1} {\rm w}\rq{}=0~
\end{equation}
is satisfied by the generalized Fresnel integrals ${\cal C}_n(z)$ and ${\cal S}_n(z)$, and so the general solution can be written as the superposition
\begin{equation}\label{eq7}
{\rm w}(z)=\alpha {\cal C}_n(z)+\beta {\cal S}_n(z)+{\rm w}(0)~.
\end{equation}
In addition, (\ref{eq1}) can be reduced to the second order ODE
\begin{equation}\label{eq2}
v''-\frac{(n-1)}{z} v'+\left(p z^{n-1}\right)^2 v =0
\end{equation}
 by using ${\rm w}\rq{}(z)=v (z)$, with $^\prime=\frac{d}{dz}$.
Letting $z^n=\zeta$, we obtain the simple harmonic oscillator equation
\begin{equation}\label{eq3}
\frac{d^2v}{d\zeta^2}+\left(\frac{p}{n}\right)^2v=0~.
\end{equation}
Thus, the solution for (\ref{eq2}) is
\begin{equation}\label{eq5}
v(z)=c_1 \cos \left( \frac{p}{n} z^{n} \right) +c_2 \sin \left( \frac{p}{n} z^{n} \right)~,
\end{equation}
and by one  integration, one obtains the solution (\ref{eq7}).

On the other hand, using the  logarithmic derivative $y(z)=\frac {v'(z)}{v(z)}$, (\ref{eq2}) becomes the Riccati equation
\begin{equation}\label{eq10}
y\rq{}+y^2=\frac{n-1}{z} y- p^2 z^{2n-2}
\end{equation}
with  particular solution
\begin{equation}\label{eq11}
y_{p}(z)=i p z^{n-1}\,.
\end{equation}

To construct the  general solution of Riccati equation (\ref{eq10}) using any particular solution  $y_p$,  we write the Bernoulli ansatz
 \begin{equation}\label{mhG}
y_g(z)=y_p(z)+\frac{1}{u(z)}~,
\end{equation}
where $u$ satisfies  the linear equation
\begin{equation}\label{mh5}
u'+\left(\frac{n-1}{z}-2 y_p\right)u=1~.
\end{equation}
The solution of (\ref{mh5}) is
\begin{equation}\label{mh6}
u(z)=\frac{\gamma+\int_0^z\mu(s)  ds}{\mu(z)}~,
\end{equation}
where $\mu(z)$ is the integrating  factor
\begin{equation}\label{mh7}
\mu(z)=z^{n-1}e^{-2i\frac{p}{n}z^n}~,
\end{equation}
which gives the general Riccati solution in the form
\begin{equation}\label{mh8}
y_g(z)=y_p(z)+\frac{\mu(z)}{\gamma+\int_0^z\mu(s) ds}~,
\end{equation}
and $\gamma$ arbitrary.
Using (\ref{mh7}) in (\ref{mh8}) and redefining the integration constant as $\gamma=\frac{i(\theta+1)}{2p}$
provides the explicit form of the general Riccati solution
\begin{equation}\label{es4}
y_g(z)=ip z^{n-1}\left(\frac{\theta e^{2i \frac{p}{n} z^n}-1}{\theta e^{2i \frac{p}{n} z^n}+1}\right)~.
\end{equation}
We then find a polar form of the general solution for ~(\ref{eq2}) using $v_g(z)=Re^{\int y_g(z)dz}$
to obtain
\begin{equation}\label{es5}
v_g(z)=R\left( e^{-i\frac {p}{ n} z^n}+\theta  e^{i\frac {p}{n} z^n}\right)~,
\end{equation}
By one integration, assuming ${\rm w}(0)=0$ and using Euler's formula, the deformed solution of (\ref{eq1}) is given by
\begin{equation}\label{es6}
{\rm w}_g(z;n)=R[(1+\theta)\, {\cal C}_n(z)+i(-1+\theta)\,{\cal S}_n(z)]
\end{equation}
and letting  $\theta=a+i b\ne 0$, we obtain
\begin{equation}\label{es8}
\begin{array}{l}
{\rm w}_{\mathcal R}(z;n)=
R\big[(a+1)\,{\cal C}_n(z)-b\,{\cal S}_n(z)\big]~, \\
{\rm w}_{\mathcal I}(z;n)=
R\big[b\,{\cal C}_n(z)+(a-1)\,{\cal S}_n(z)\big].
\end{array}
\end{equation}
Comparing (\ref{eq7}) with (\ref{es8}), one can see that we managed to replace the superposition constants $\alpha$ and $\beta$ by the real and imaginary components of the parameter entering the general Riccati solution. This replacement helps to disentangle an underlying supersymmetric structure of the solution expressed in this way as it will be shown in the next section.

\begin{center}
\begin{figure}
\includegraphics[width=1.05\textwidth]{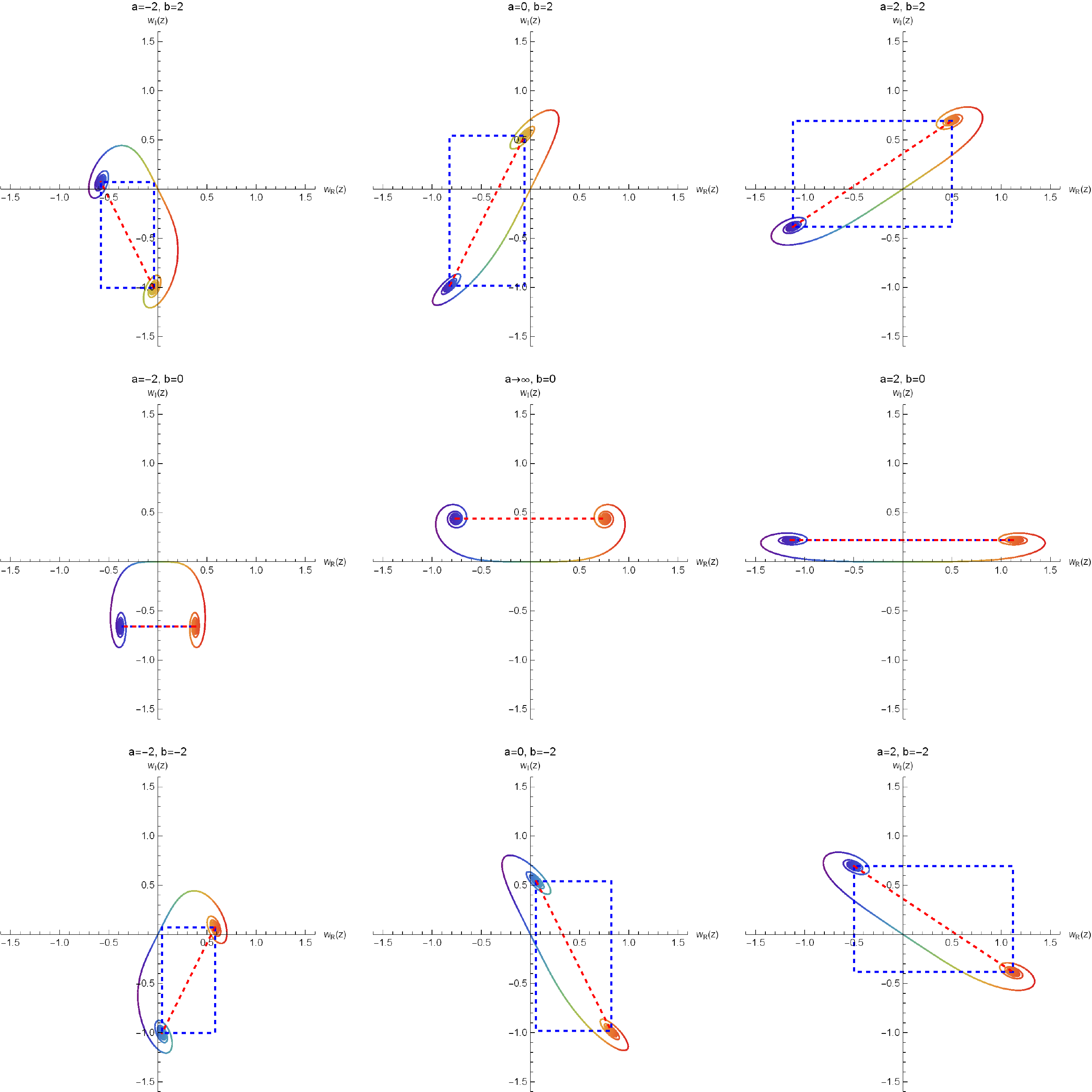}
\caption{The generalized ${\cal F}_3$ spiral for different values of the parameters $a$ and $b$.}
\label{fig1}
\end{figure}
\end{center}

Argand plots of the generalized Fresnel spirals for $n=3$ are shown in Fig.~\ref{fig1}
based on different values of the deformation parameter $\gamma =\frac{-b+i(a+1)}{2 \pi}$. In the center, when $a=\infty$ and $b=0$, we used the standard undeformed  spiral given by ${\rm w}_{\mathcal R}(z;3)={\cal C}_3(z)$ and ${\rm w}_{\mathcal I}(z;3)={\cal S}_3(z)$. All figures, except the $a=\infty,~b=0$ case, are scaled by the factor $R=1/\sqrt{a^2+b^2}$.

To find the analytical expressions of the generalized Fresnel integrals for $n\ge 2$, together with their foci corresponding to $z\rightarrow \pm\infty$, we proceed as follows.
First, we let $\varphi(s)=\frac{p}{n}s^n$, and we notice that $d \varphi=\frac{ds}{\rho(s)}$, and by inverting we find $\rho(\varphi)=\frac{n^{\frac{1-n}{n}}}{p^{\frac 1 n}}\varphi^{\frac{1-n}{n}}$, thus \eqref{Fes1} and \eqref{Fes2} become
\begin{eqnarray}
	{\cal C}_n(z^{\star}) &=&\frac{n^{\frac{1-n}{n}}}{p^{\frac 1 n}} \int_{0}^{z^{\star}} \varphi^{\frac{1-n}{n}}\cos  \varphi ~d \varphi ~, \label{Fes3}\\
	{\cal S}_n(z^{\star})&=&\frac{n^{\frac{1-n}{n}}}{p^{\frac 1 n}} \int_{0}^{z^{\star}}\varphi^{\frac{1-n}{n}}\sin  \varphi ~d \varphi ~, \label{Fes4}
\end{eqnarray}
with the upper limit given by  $z^{\star}=\frac{p}{n}z^n$.
These integrals can now be found in closed form in terms of the hypergeometric function, as they are
\begin{eqnarray}
	{\cal C}_n(z^{\star}) &=&\left(\frac{n z^{\star}}{p}\right)^{\frac 1 n}{}_1F_{2}\left[\frac{1}{2n};\frac 1 2,1+\frac{1}{2n};-\left(\frac{z^{\star}}{2}\right)^2\right]~, \label{Fes5}\\
	{\cal S}_n(z^{\star})&=& \frac{z^{\star}}{1+n}\left(\frac{n z^{\star}}{p}\right)^{\frac 1 n} \label{Fes6}\\&&{}_1F_{2}\left[\frac 1 2 \left(1+\frac 1 n\right);\frac 3 2 , \frac 1 2 \left(3+\frac 1 n\right);-\left(\frac{z^{\star}}{2}\right)^2\right]~ \notag.
\end{eqnarray}
The location of the foci of the generalized undeformed spirals are obtained from \eqref{Fes5} and \eqref{Fes6} by letting $z^\star \rightarrow \pm\infty$. The focus in the first quadrant has the coordinates given by
\begin{eqnarray}
x_{_{{\cal F}_n}}&=&\left(\frac{n}{p}\right)^{\frac{1}{n}}\Gamma\left(1+\frac{1}{n}\right)\cos\left(\frac{\pi}{2n}\right)~, \label{xf}\\
y_{_{{\cal F}_n}}&=&\left(\frac{n}{p}\right)^{\frac{1}{n}}\Gamma\left(1+\frac{1}{n}\right)\sin\left(\frac{\pi}{2n}\right)~, \label{yf}
\end{eqnarray}
whereas the second focus can be found in the second quadrant at $(-x_{_{{\cal F}_n}},\,y_{_{{\cal F}_n}})$ if $n$ is odd and in the third
quadrant at $(-x_{_{{\cal F}_n}},\,-y_{_{{\cal F}_n}})$ if $n$ is even.
Both foci lie on the circle of radius
$$
r_n=\left(\frac{n}{p}\right)^{\frac{1}{n}}\frac{1}{n}\Gamma\left(\frac{1}{n}\right)~.
$$
Moreover, for $n\gg 1$, we use the expansion  
$$
\Gamma\left(\frac 1 n \right)\approx n-\gamma+\frac{1}{2}\left(\gamma^2+\frac{\pi^2}{6}\right)\frac{1}{n}+O\left(\frac{1}{n^2}\right)~,
$$
where $\gamma=0.577$ is the Euler-Mascheroni constant, and the limit
$\left(n^{1/n}\right)_{n\rightarrow\infty}=1$, to show
that $y_{_{{\cal F}_{\infty}}}\rightarrow 0$ and $x_{_{{\cal F}_{\infty}}}\rightarrow 1$.
Consequently, at ever increasing $n$, these type of spirals are stretched more and more along the [-1,1] segment
of the abscissa axis with the two foci at the end points. Concerning the deformed spirals, the coordinates of the foci are related to the undeformed foci through (\ref{es8})
\begin{eqnarray}\label{deff}
x_{_{{\cal F}_n}}^d&=&R\big[(a+1)\,x_{_{{\cal F}_n}}-b\,y_{_{{\cal F}_n}}\big]~,\label{xfd}\\
y_{_{{\cal F}_n}}^d&=&R\big[b\,x_{_{{\cal F}_n}}+(a-1)\,y_{_{{\cal F}_n}}\big]~,\label{yfd}
\end{eqnarray}
or in matrix form
\begin{equation}\label{mform}
\left(\begin{array}{c}
x_{_{{\cal F}_n}}^d\\ y_{_{{\cal F}_n}}^d \end{array}\right)=\left( \begin{array}{cc}
\cos \delta & -\sin \delta  \\
\sin \delta & \cos \delta
\end{array} \right)\left(\begin{array}{c}
x_{_{{\cal F}_n}}\\ y_{_{{\cal F}_n}} \end{array}\right)+\frac{1}{\sqrt{a^2 +b^2}}\sigma_3\left(\begin{array}{c}
x_{_{{\cal F}_n}}\\ y_{_{{\cal F}_n}} \end{array}\right)
\end{equation}
with $\cos\delta=a/\sqrt{a^2+b^2}$, $\sin\delta=b/\sqrt{a^2+b^2}$, and $\sigma_3$
the Pauli $z$-spin-flip matrix showing the rotation part and the flip part of the deformation.

To illustrate these results, we choose the odd case $n=3$. Using the value of $z^{\star}$ for $n=3$ and $p=\pi$, \eqref{Fes5} and \eqref{Fes6} are
\begin{eqnarray}
	{\cal C}_3(z) &=&z~{}_1F_{2}\left[\frac 1 6 ; \frac 1 2 , \frac 7 6 ;-\left(\frac{\pi z^{3}}{6}\right)^2\right]~, \label{Fes7}\\
	{\cal S}_3(z)&=& \frac{\pi}{12} z^4~{}_1F_{2}\left[\frac 2 3 ; \frac 3 2 , \frac 5 3 ;-\left(\frac{\pi z^{3}}{6}\right)^2\right]\label{Fes8}
\end{eqnarray}
with the following coordinates of the foci
\begin{eqnarray}
x_{_{{\cal F}_3}}&=&\pm\frac{\sqrt{3}}{6} \left(\frac{3}{\pi}\right)^{\frac 1 3}\Gamma\left(\frac 1 3 \right)\approx \pm 0.762~,\\
y_{_{{\cal F}_3}}&=&\frac 1 6 \left(\frac{3}{\pi}\right)^{\frac 1 3}\Gamma\left(\frac 1 3 \right)\approx 0.440~.
\end{eqnarray}

\begin{center}
\begin{figure}
\includegraphics[width=0.95\textwidth]{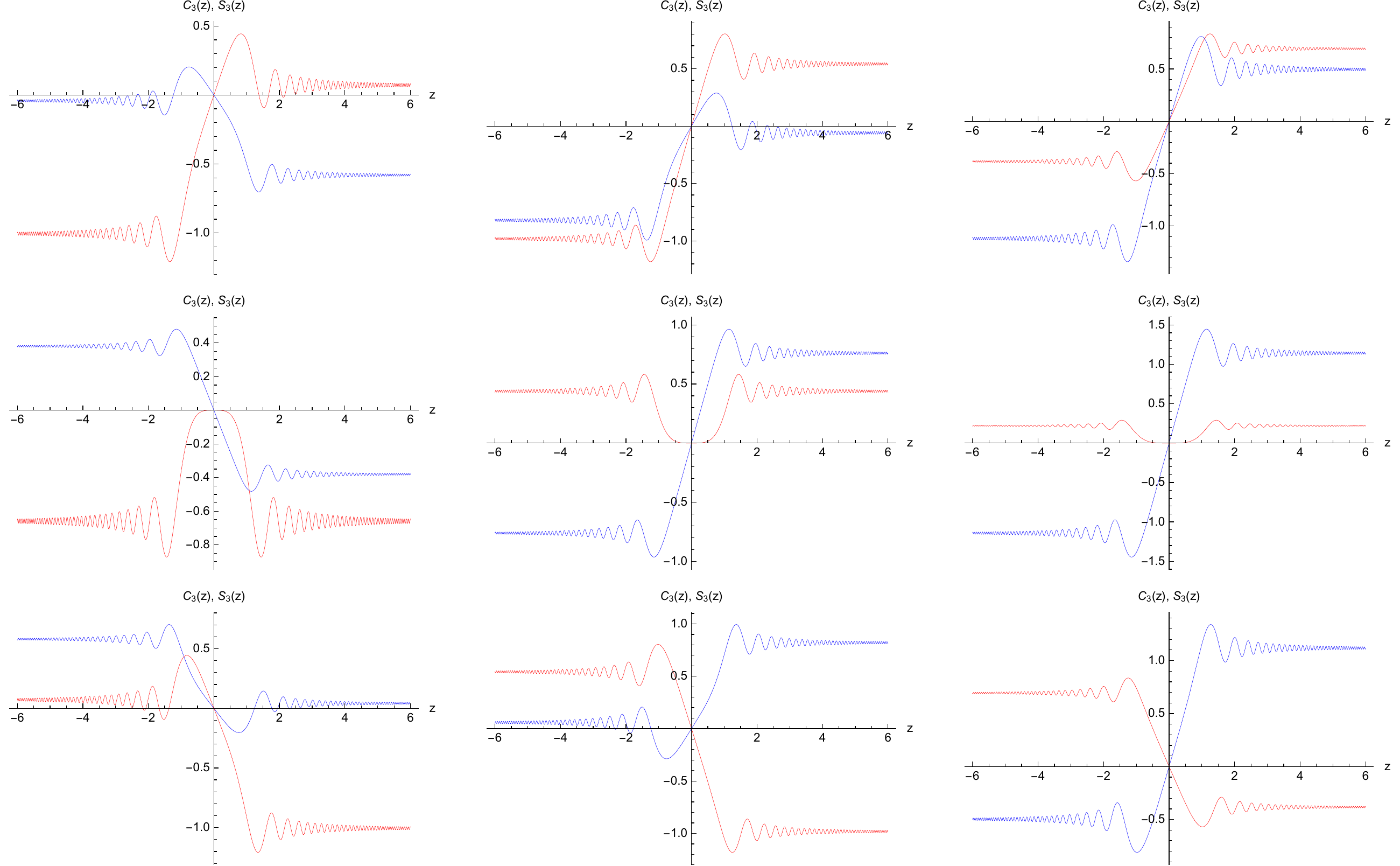}
\caption{Generalized Fresnel integrals for $n=3$, for the same values of the parameters $a$ and $b$ given in Fig. \ref{fig1}.}
\label{fig2}
\end{figure}
\end{center}

 In Fig.~\ref{fig2},  we show the graphs of the generalized Fresnel integrals for $n=3$ for the same values of parameters $a$ and $b$ from Fig.~\ref{fig1}. The location of the horizontal asymptotes are the foci for $z>0$ for the center plot when $a=\infty , b=0$ which corresponds to the standard spiral.
\section{The supersymmetric approach}
There are hidden supersymmetric features in the deformations of these spirals generated by the complex Riccati parameter. To reveal the supersymmetric aspects, we resort on the factorization technique for ordinary differential equations, which has been used by many authors to solve quantum mechanical eigenvalue problems and in the study of isospectral problems in the area of supersymmetric quantum mechanics, for reviews see \cite{s3,Dong}. Thus we write
equation~(\ref{eq2}) in the factorized form $A^-A^+v=0$ using the factoring operators given by
\begin{equation}
\begin{array}{l}
A^+=z^{-\frac{n-1}{2}}\frac{d}{dz}+pz^{\frac{n-1}{2}}\tan \frac{p z^n}{n}\\
\\
A^-=\frac{d}{dz}z^{-\frac{n-1}{2}}-pz^{\frac{n-1}{2}}\tan \frac{p z^n}{n}~.
\end{array}
\end{equation}

\begin{center}
\begin{figure}
\includegraphics[width=0.95\textwidth]{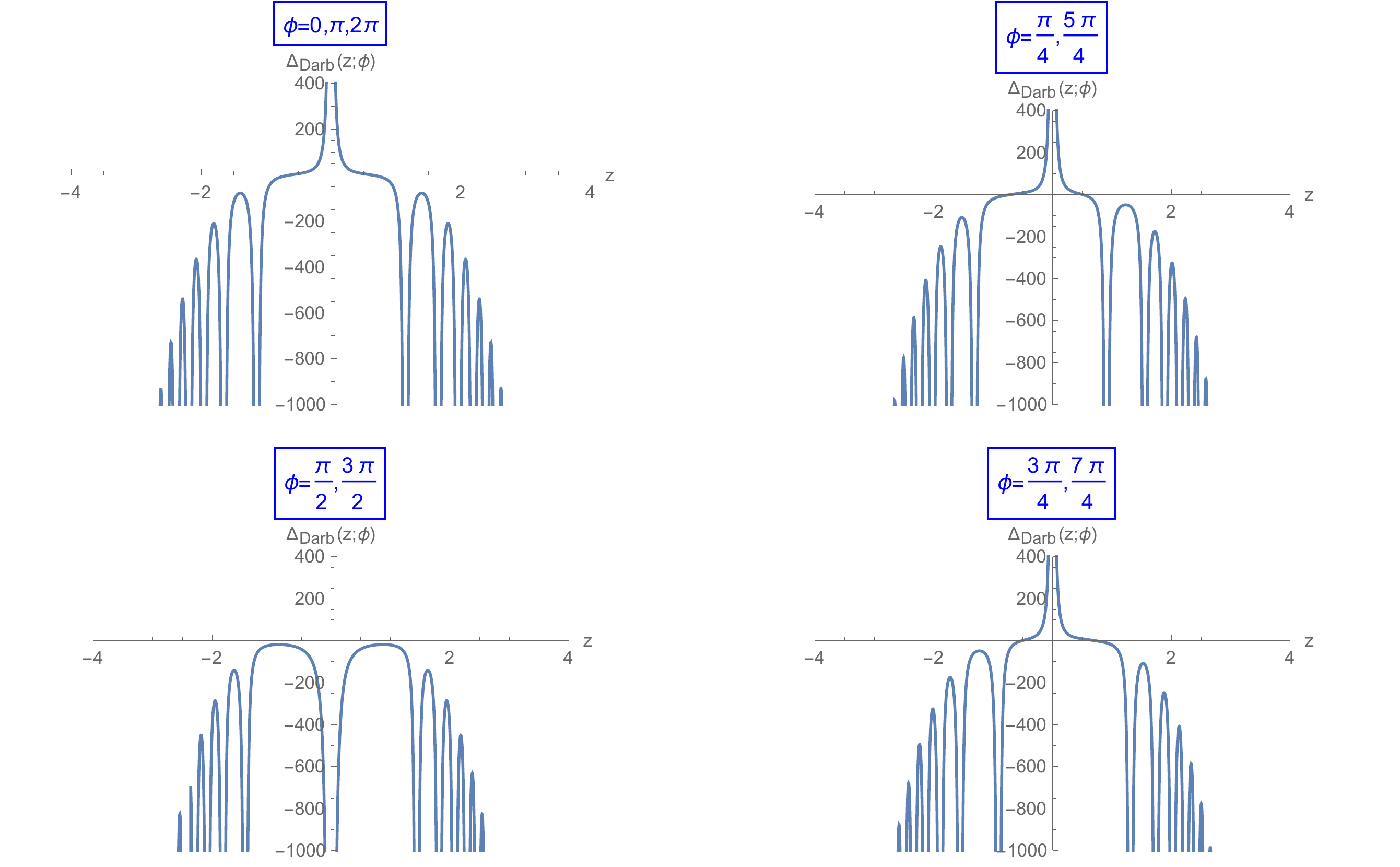}
\caption{The Darboux distortion of \eqref{eq2} for various phases of $\phi$. }
\label{fig3}
\end{figure}
\end{center}

To find out the explicit second-order linear ODE that corresponds to the generalized deformed Cornu spirals, we first write the general Riccati solution (\ref{es4}) in the trigonometric form
\begin{equation}\label{es5b}
y_g(z)=-p z^{n-1}\tan \left (\frac{p z^n}{n}+\phi \right)~, \qquad \theta =R e^{i\phi}~,
\end{equation}
and like in supersymmetric quantum mechanics we employ it in the supersymmetric partner equation $A^+A^-\widetilde{V}=0$, i.e.,
\begin{equation}\label{factgen}
\left[z^{-\frac{n-1}{2}}\frac{d}{dz}-z^{-\frac{n-1}{2}}y_g(z)
\right]
\left[\frac{d}{dz}z^{-\frac{n-1}{2}}+z^{-\frac{n-1}{2}}y_g(z)
\right]\widetilde{V}=0~. 
\end{equation}
We then obtain this equation as
\begin{equation}\label{feq3bb}
\widetilde{V}\rq{}\rq{}-\frac{n-1}{z} \widetilde{V} \rq{}+\left[\left(p z^{n-1}\right)^2+\Delta_{{\rm Darb}}(z;n,\phi)\right]\widetilde{V} =0~,
\end{equation}
with the Darboux distortion, which depends parametrically on the phase shift $\phi$, given by
\begin{eqnarray}\label{Ddp}
&\Delta_{{\rm Darb}}(z;n,\phi)=-2\left(p z^{n-1}\right)^2+\frac{n^2-1}{4z^2}\nonumber\\
&-(n-1)pz^{n-2} \tan  \left (\frac{p z^n}{n}+\phi \right)\\
& -2\left(p z^{n-1}\right)^2\tan ^2 \left (\frac{p z^n}{n}+\phi \right)~.\, \nonumber
\end{eqnarray}
In Fig.~\ref{fig3}, we display various cases of the parametric Darboux distortions $\Delta_{{\rm Darb}}(z;\phi)$ of the deformed Cornu spirals
presented in Fig.~\ref{fig1}. We notice the negative parabolic envelope as given by the first term in (\ref{Ddp}) together with the singularities due to the terms containing the tangents for nonzero $z$. The singularities at the origin are due to the $1/z^2$ term except for the cases $\phi =\pi/2$ and $3\pi/2$ when the dominant contribution comes from the cotangent terms.  For these values of phase, the Darboux distortion simplifies to
\begin{equation}\label{darb}
\begin{array}{ll}
&\Delta_{{\rm Darb}}(z;0,\pi)=\frac{2}{z^2}- 2\pi^2 z^4 
\frac{\left[1 +\frac{1}{3}{\rm sinc}\left(\frac {2\pi z^3}{3}\right) \right]}{\cos^2\left(\frac {\pi z^3}{3}\right)}~,\nonumber \\
&\Delta_{{\rm Darb}}(z;\frac{\pi}{4},\frac{ 5 \pi}{4})=\frac{2}{z^2}-\frac{2\pi z\left[2 \pi z^3+\cos \left(\frac{2\pi z^3}{3}\right)\right]}{1-\sin \left(\frac{2\pi z^3}{3}\right)}~,\nonumber \\
&\Delta_{{\rm Darb}}(z;\frac{\pi}{2}, \frac{3\pi}{2})=\frac{2}{z^2}-2\pi^2 z^4
\frac{\left[1 -\frac{1}{3}{\rm sinc}\left(\frac {2\pi z^3}{3}\right) \right]}{\sin^2\left(\frac {\pi z^3}{3}\right)}~,\nonumber \\
&\Delta_{{\rm Darb}}(z;\frac{3\pi}{4}, \frac{7 \pi}{4})=\frac{2}{z^2}-\frac{2\pi z\left[2 \pi z^3-\cos \left(\frac{2\pi z^3}{3}\right)\right]}{1+\sin \left(\frac{2\pi z^3}{3}\right)}~.
\end{array}
\end{equation} 
The general solution to (\ref{feq3bb}) can be found by letting $A^-\widetilde V=\widetilde  U_h$, and first solving the
homogenous equation
$A^+\widetilde U=0$ with solution
\begin{equation}\label{phih}
\widetilde U_h(z;n, \phi)=b_1 \cos  \left (\frac{p z^n}{n}+\phi \right).
\end{equation}\label{psih}
We then obtain the general solution $\widetilde V$ 
in the form
\begin{eqnarray}\label{psig}
&\widetilde V(z;n,\phi)=\frac{1}{4n} 
~ z^{\frac{n-1}{2}}\sec \left (\frac{p z^n}{n}+\phi \right)\nonumber\\
&\Bigg[2n 
\left(b_1 z \cos \phi +2 b_2\right)- \Bigg. \nonumber \\
& \Bigg. b_1 z \left(e^{i \phi}E_{\frac{n-1}{n}}\left(-\frac{2 i p z^n}{n}\right)+e^{-i \phi}E_{\frac{n-1}{n}}\left(\frac{2 i p z^n}{n}\right)\right)\Bigg],
\end{eqnarray}
where $E_m(\tau)$ is the generalized exponential integral function given by
$E_m(\tau)=\int_1^\infty \frac{e^{-\tau s}}{s^m}~ds$.

\section{Conclusion}

The spirals ${\cal F}_n$, of which the particular case ${\cal F}_2$ is the Cornu spiral, have been defined in this Letter, and their parametric Darboux deformations generated through the corresponding general Riccati solution have been studied in detail, with focus on the odd case ${\cal F}_3$. Geometrically, these deformations are generated by the two independent shifts, $a$ and $b$, along the two orthogonal axes of the plane in which the spirals are plotted. These shifts determine both the deformation of the rolls of the spirals and its global rotation as seen in the plots. One can think of possible applications similar to those of the Cornu (clothoid) spiral, such as higher-order optical diffraction and transition curves in road and railway alignments.

\acknowledgments{We thank the referees for their careful reading and thoughtful suggestions.}

\bigskip

\newpage

\end{document}